\documentclass[journal,twoside,web]{ieeecolor}
\usepackage{tmi}
\usepackage{cite}
\usepackage{amsmath,amssymb,amsfonts}
\usepackage{algorithmic}
\usepackage{graphicx}
\usepackage{textcomp}

\def\BibTeX{{\rm B\kern-.05em{\sc i\kern-.025em b}\kern-.08em
    T\kern-.1667em\lower.7ex\hbox{E}\kern-.125emX}}
\begin{document}
\title{Deep Multi-Resolution Dictionary Learning for Histopathology Image Analysis}
\author{Nima Hatami, Mohsin Bilal, and Nasir Rajpoot, \IEEEmembership{Senior Member, IEEE}

\thanks{This work was supported by the UK Medical Research Council (award MR/P015476/1). NR is also supported by the PathLAKE digital pathology consortium, which is funded from the Data to Early Diagnosis and Precision Medicine strand of the government’s Industrial Strategy Challenge Fund, managed and delivered by UK Research and Innovation (UKRI).}
\thanks{N. Hatami is with Computer Science Department, University of Warwick, UK, and Tissue Image Analytics (TIA) Centre, University of Warwick, UK (e-mail: Nima.Hatami@warwick.ac.uk).}
\thanks{{M. Bilal is with Computer Science Department, University of Warwick, UK, and Tissue Image Analytics (TIA) Centre, University of Warwick, UK (e-mail: Mohsin.Bilal@warwick.ac.uk).}}
\thanks{N. Rajpoot is with Computer Science Department, University of Warwick, UK, and Tissue Image Analytics (TIA) Centre, University of Warwick, UK, and Department of Pathology, University Hostpials Coventry \& Warwickshire, UK (e-mail: N.M.Rajpoot@warwick.ac.uk).}}

\maketitle

\begin{abstract}
The problem of recognizing various types of tissues present in multi-gigapixel histology images is an important fundamental pre-requisite for downstream analysis of the tumor microenvironment in a bottom-up analysis paradigm for computational pathology. In this paper, we propose a deep dictionary learning approach to solve the problem of tissue phenotyping in histology images. %In this paper, histopathology image classification is treated as texture recognition and application of deep dictionary learning is investigated in computational pathology domain.
We propose deep Multi-Resolution Dictionary Learning (deep MRDL) in order to benefit from deep texture descriptors at multiple different spatial resolutions. We show the efficacy of the proposed approach through extensive experiments on four benchmark histology image datasets from different organs (colorectal cancer, breast cancer and breast lymph nodes) and tasks (namely, cancer grading, tissue phenotyping, tumor detection and tissue type classification). We also show that the proposed framework can employ most off-the-shelf CNNs models to generate effective deep texture descriptors. 
%Performance of the proposed deep MRDL is compared to the standard deep learning algorithms and state-of-the art results for each task. Accuracy improvements obtained on CRC-TIA, PCam, NCT-CRC, and BreakHis datasets, and also qualitative evaluation is carried out on TCGA diagnostic slides. 
\end{abstract}

\begin{IEEEkeywords}
Computational Pathology, Colorectal Cancer Grading, Dictionary Learning, Deep Learning, H\&E Staining, Tissue Classification, Texture Recognition.
\end{IEEEkeywords}

\section{Introduction}

\IEEEPARstart{W}{ith} a growing number of cancer cases worldwide, workload of expert histopathologists is increasing dramatically \cite{azam2020diagnostic}. A 2018 survey by the Royal College of Pathologists reported acute shortage of histopathologists in the UK with the problem likely to grow further in the coming years. Digitalization of histopathology slides has led to computational pathology playing an important role by using the power of deep learning, image analytics and big data \cite{snead2016validation}. This not only helps to assist with the diagnostic process resulting in efficiency gains \cite{griffin2017digital,colling2019artificial} but also to enhance diagnostic precision \cite{niazi2019digital,bera2019artificial}. 

Availability of large volumes of histology image data coupled with advances in computational power have given rise to the development of novel deep learning methods to effectively tackle many pathology tasks. Successful organization of recent challenge competitions have shown that it is possible to outperform the average expert pathologist's performance leading to the prospect of computer-assisted provision of healthcare \cite{bejnordi2017diagnostic}. However, there are several challenges that have arisen in the relatively new field of computational pathology. One of the main challenges in order to use deep learning in a routine clinical setting is the lack of high accuracy with high consistency. In fact, there are already many algorithms for different computational pathology tasks such as cell classification, gland segmentation, tissue phenotyping, tumor detection that have been shown to work well in a laboratory settings, but most are not ready to cope with `pathology image data from the wild'.
There are a few methods that are trained and validated on multiple cohorts. Despite scientifically encouraging results, yet far from being ideal for the real-world challenges \cite{kather2020pan, kather2019predicting, kather2019deep, fu2020pan}. And so the quest for efficient and effective models for histopathology images that are clinical grade is ongoing and is active area of research for the medical imaging community.

\begin{figure*}[!t]
\includegraphics[width=\textwidth]{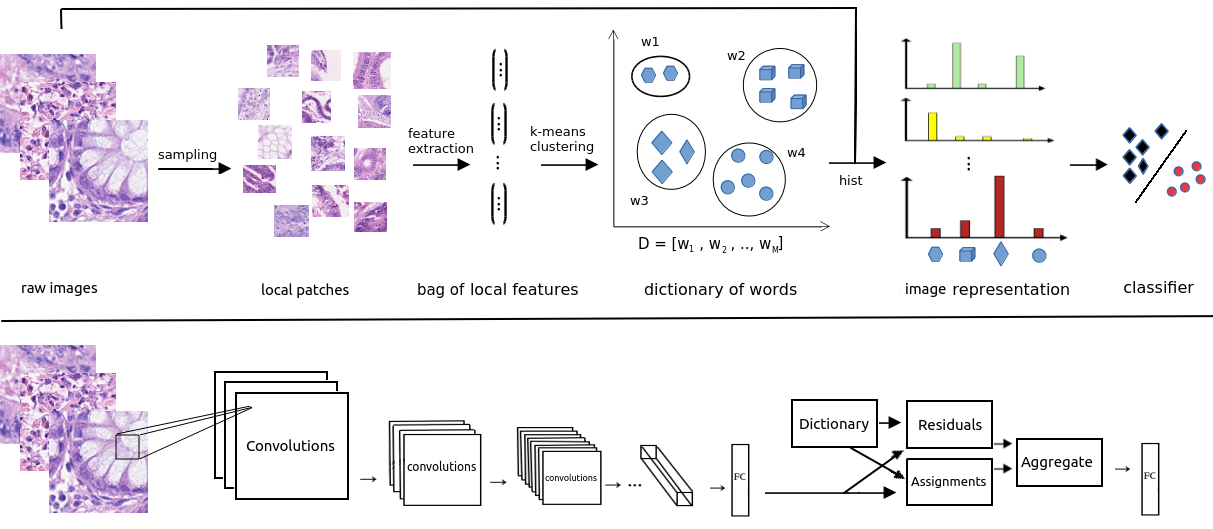}
\caption{Traditional BoW model (top) vs. deep dictionary learning network with end-to-end learning framework (bottom) for histopathology image classification.} 
\label{fig_bow_vs_DL}
\end{figure*}

Texture is ubiquitous and provides useful cues of material properties of objects and their identity. Visual representations based on orderless aggregations of local features (e.g. SIFT \cite{Lowe04}, SURF \cite{SURF}, HOG \cite{HOG}, and LBP \cite{LBP}) were originally developed as image texture descriptors. Before the `deep learning era', models such as the histograms of filter bank responses \cite{Leung01, Cula01}, Bag-of-visual-Words (BoW) \cite{Csurka04} and Fisher vector \cite{Perronnin07} have been successfully applied to texture recognition. The main idea of BoW is to represent an image by a histogram of feature words of a data-learned dictionary (the so-called {\em codebook}). The BoW model usually consists of the following five steps: i) creating local patches from the input images, ii) extracting local features from the patches and creating the `bag', iii) learning a dictionary of visual words, iv) quantizing features using dictionary and representing samples by histogram of words (encoding), and v) classification of features using SVM or decision tree. A block diagram of the traditional BoW is shown in Figure \ref{fig_bow_vs_DL} ({\em top}). In fact, convolutional neural networks (CNNs) can be viewed as BoW models too. The convolution layer of CNNs operates in a sliding window manner acting as a local feature extractor. The output feature maps preserve a relative spatial arrangement of input images. The resulting globally ordered features are then concatenated and fed into a fully connected (FC) layer which acts as a classifier. Despite massive achievement on a broad range of applications such as image classification, object detection, segmentation, and scene understanding, CNNs are typically not ideal for texture recognition due to the need for a spatially invariant representation describing the feature distributions instead of concatenation.

With the rapid growth of deep learning, handcrafted features and predefined filter banks in the BoW paradigm were replaced by the `deep features' and have achieved state of-the-art results \cite{Cimpoi15, cimpoi2016deep}. The main challenge faced by these methods is that each component is optimized in a separate step, and features and encoders are fixed once built. Therefore, feature learning (CNNs and dictionary) does not benefit from labelled data. Recently, a deep dictionary learning approach has been proposed for texture recognition \cite{Zhang17}. It outperforms existing modular methods and achieves the state-of-the-art results on material/texture datasets. Also, it is an end-to-end learning framework whereby features, dictionaries, encoding representation and the classifier are all learned simultaneously facilitated by a suitable loss function. 

%From computer vision and image analysis perspectives, histopathology images analysis tasks such as tumor detection, cancer grade, and tissue phenotyping can be treated as material and texture recognition problem. 
In this paper, we propose a novel deep dictionary learning method using a CNN backbone and investigate its effectiveness for tissue phenotyping casted as a texture recognition problem. Our main contributions are listed below:
\begin{enumerate}
    \item We propose a deep multi-resolution dictionary learning (MRDL) framework building multiple dictionaries on different layers of the network in order to benefit from different scales of descriptors.
    \item We show that the fusion of information coming from different dictionaries built at multiple resolutions helps the entire model to learn a better context of the input image, thus boosting the overall accuracy.
    \item Through extensive experiments on various H\&E-stained image datasets of different organs and pathology tasks, we demonstrate the robustness of the proposed deep MRDL as compared to several other state-of-the-art methods.
    \item Finally, we provide solid mathematical formulations for the proposed MRDL that demonstrate end-to-end training and optimization of the network. This helps better understating of the MRDL model, and its effectiveness.
\end{enumerate}

The rest of the paper is organized as follows: a review of recent work on histology image classification is presented in the next section. The proposed deep MRDL framework is described in detail in section \ref{Methodology}. A description of datasets used in this study and experimental results are reported in section \ref{Experiments}. Finally, section \ref{Conclusions} concludes the paper and outlines the future directions. 

%\hfill \date{\today}

\section{Related Work}
\label{Related_work}

This section briefly reviews the recent research on histology tissue image classification. More specifically, we focus on deep learning methods for the classification of H\&E images.  

%Kather {\em et al.} \cite{kather2016multi} proposed to use six different types of traditional texture descriptors such as LBP, gabor filters, lower-order and higher-order histogram features for the classification of eight different tissue phenotypes in colon cancer histology images. Although, the discrimination performance improved, the texture descriptors do not fully capture the biological significance of the tissue components, hence this method is not very accurate in identifying tumors with complex stroma and mucosa.
%

Kather {\em et al.} \cite{kather2019predicting} proposed a two-step deep learning model to predict microsatellite instability (MSI) directly from histology in gastrointestinal cancer. They first detect the tumor regions in a H\&E image. Then, train a second CNNs to classify the tumor patches into MSI and MSS. The experimental results on the TCGA data shows the potential of the proposed image-based method against the expensive alternative of the gene expression-based approaches for MSI detection.  

Fu {\em et al.} \cite{fu2020pan} recently proposed pan-cancer computational histopathology model for predicting mutations, tumor composition and prognosis. They classified the tissue origin across organ sites and provides highly accurate, spatially resolved tumor and normal distinction within a given slide. Afterwards, they used CNNs to extract histopathological patterns and accurately discriminates 28 cancer and 14 normal tissue types. Furthermore, they predicted whole genome duplications, focal amplifications and
deletions, as well as driver gene mutations. Experimental results demonstrate wide-spread correlations with gene expression indicative of immune infiltration and proliferation. 

Pan-cancer image-based detection of clinically actionable genetic alterations proposed by Kather {\em et al.} \cite{kather2020pan}. The idea is to take CNNs pre-trained on ImageNet, and apply transfer learning on the tumor patches. They showed that deep learning can predict point mutations, molecular tumor subtypes and immune-related gene expression signatures, directly from routine histological images of tumor tissue.

Utilizing recent findings on rotation equivariant CNNs, Veeling {\em et al.} \cite{PCam} proposed a model that leverages rotational symmetries in a principled manner. They presented a visual analysis showing improved stability on predictions, and demonstrate that exploiting rotation equivariance significantly improves tumor detection performance on a challenging lymph node metastases dataset. 

Tellez {\em et al.} \cite{tellez2019neural} proposed Neural Image Compression (NIC), a two-step method to build convolutional neural networks for gigapixel image analysis solely using weak image-level labels. First, gigapixel images are compressed using a neural network trained in an
unsupervised fashion, retaining high-level information while suppressing pixel-level noise. Second, a CNN is trained on these compressed image representations to predict image-level labels, avoiding the need for fine-grained manual annotations.

A GAN based semi-supervised learning (SSL-GAN) pipeline is proposed to use most of the unlabeled data available in order to target the detection of oral cancer metastasis in cervical lymph nodes \cite{Navid_miccai20}. Due to a large number of images and the need for expert knowledge, collecting annotations for this task is time-consuming and labor-intensive. They showed that not only can the proposed SSL-GAN improve the classification with a relatively modest budget of annotations, it also performs well in transferring representation across different domains for pathology images.

Hatami {\em et al.} \cite{DeepFEL} propose Deep Fastfood Ensembles – a simple, fastand yet effective method for combining deep features pooled from popular CNN models pre-trained on totally different source domains (e.g.,natural image objects) and projected onto diverse dimensions using random projections, the so-called Fastfood \cite{Fastfood}. The final ensemble output is obtained by a consensus of simple individual classifiers, each of which is trained on a different collection of random basis vectors. This offers extremely fast and yet effective solution for tissue image classification, especially when training times and domain labels are of the essence.

Recently, Shaban {\em et al.} \cite{Shaban20} presented a context-aware deep neural network which is able to incorporate larger context than standard CNN based patch classifiers. It first encodes the local representation of a histology image into high dimensional features, then aggregates the features by considering their spatial organization to make a final prediction. 

A histological visual dictionary is generated for predicting prognosis in intrahepatic cholangiocarcinoma (ICC) \cite{muhammad2019towards}. They propose an unsupervised deep convolutional autoencoder-based clustering model that groups together cellular and structural morphologies of tumor in 246 WSI, based on visual similarity. From this visual dictionary of histologic patterns, the clusters were used as covariates to train Cox-proportional hazard survival models. 

Recently, some graph-based approaches have also been proposed for the problem of histology image classification. 
Javed {\em et al.} \cite{javed2020cellular} proposed a semi-supervised cellular community detection algorithm for tissue phenotyping based on cell detection and classification, and clustering of image patches into biologically meaningful communities. First a CNN is used for cell detection and classification and then based on potential cell-cell connections between these cells, feature vectors are computed at the patch level. These feature vectors are then used to construct an image level network using chi-square distance such that each node is a patch in WSI and edges have weights inversely proportional to the distance between the feature vectors. In this network, geodesic distances are computed which are then used to compute node clusters such that each cluster corresponds to a particular tissue phenotype. In order to incorporate the entire tissue micro-architecture for CRC grading purposes, Zhou {\em et al.} \cite{Zhou19} proposed cell-graph convolutional neural network (CGC-Net). It converts each large histology image into a graph, where each node is represented by a nucleus within the original image and cellular interactions are denoted as edges between these nodes according to node similarity. 
%
%Two comments about the above review of existing literature: (1) it provides a summary of every method, without really saying much about their limitations and (2) it can benefit from a structure (eg, CNNs for patch classification, graph based methods and dictionary based methods).

%{\color{red} In summary, approaches such as graph-based models are computationally expensive and require larger resources of GPU and memories. On the other hand, the classical standard CNNs-based models demand larger amount of labeled data. The proposed MRDL stands in between and is a good trade-off between computational cost and amount of data required for a good accuracy.}

\section{The Proposed Method}
\label{Methodology}

\begin{figure*}[!t]
\includegraphics[width=\textwidth]{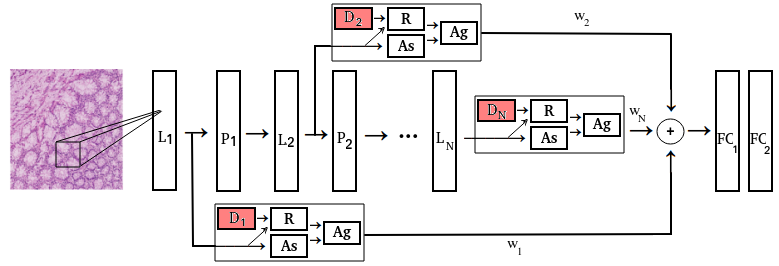}
\caption{The general block diagram of the proposed multi-resolution deep dictionary learning for any desired CNN backbone. A dictionary is building for each layer using descriptor features from that specific layer. $L_{i}, P_{i}, D_{i}, R, As, Ag, FC$ are $i$-th convolutional layers, pooling, dictionary, residual, assignment, aggregation, and fully-connected, respectively. Contribution of each dictionary is determined by learning the adaptive weights ($\omega_{i}$).} 
\label{fig_MRDL}
\end{figure*}

The proposed deep Multi-resolution Dictionary Learning (deep MRDL) model is an extension of the work by Zhang {\em et al.} \cite{Zhang17} and is built on top of the Deep TEN: Texture Encoding Network. It uses the "Encoding-Layer" (gray boxes in Figure \ref{fig_MRDL}) which consists of four blocks: dictionary, residuals, assignment weights, and aggregation. It is a novel layer integrated on top of any desired convolutional layers, which ports the entire dictionary learning and encoding pipeline into a single model. The main advantage of deep TEN compared to the previous deep dictionary learning approaches is that it is an end-to-end learning framework, where the inherent visual vocabularies are learned directly from the loss function. The features, dictionaries, encoding representation and the classifier are all learned simultaneously (gradient information passing to each component during back propagation, tuning each component). The Encoding Layer has three main properties. (1) The representation is orderless and describes the feature distribution, which is suitable for material and texture recognition. (2) The Encoding Layer acts as a pooling layer integrated on top of convolutional layers, accepting arbitrary input sizes and providing output as a fixed-length representation. By allowing arbitrary size images, the Encoding Layer makes the deep learning framework more flexible. And, (3) the Encoding Layer learns an inherent dictionary and the encoding representation which is likely to carry domain-specific information and therefore is suitable for transferring pre-trained features.

Given a set of $N$ visual descriptors $X ={x_{1},...,x_{N}}$ and a learned codebook $C ={x_{1},...,x_{K}}$ containing $K$ codewords that are $D$-dimensional, each descriptor $x_{i}$ can be assigned with a weight $a_{ik}$ to each codeword $c_{k}$ and the corresponding residual vector is denoted by $r_{ik} = x_{i} - c_{k}$, where $i=1...N$ and $k=1,...K$. Given the assignments and the residual vector, the residual encoding model applies an aggregation operation for every single codeword $c_{k}$. The resulting encoder outputs a fixed length representation ($K$ independent from the number of input descriptors $N$) $E = {\epsilon_{1},...\epsilon_{K}}$:

\begin{equation}
\label{eq_1}
    \epsilon_{k} = \sum_{i=1}^{N} \epsilon_{ik} = \sum_{i=1}^{N} a_{ik} r_{ik}
\end{equation}

Consider the assigning weights for assigning the descriptors to the codewords. Hard-assignment provides a single non-zero assigning weight for each descriptor $x_i$, which
corresponds to the nearest codeword. The k-th element
of the assigning vector is given by $a_{ik} = \mathbb{I} (\parallel r_{ik}\parallel ^2 = min\{\parallel r_{i1}\parallel ^2,...\parallel r_{iK}\parallel ^2\})$, where $\mathbb{I}$ is the indicator function (outputs 0 or 1). However, hard-assignment does not take into account the codeword ambiguity and also makes the model non-differentiable. Soft-weight assignment addresses this issue by assigning a descriptor to each codeword \cite{van2008kernel}. The assigning weight is given by:

\begin{equation}
    a_{ik} = \frac{e^{(-\beta \parallel r_{ik} \parallel ^2)}}{\sum_{j=1}^{K} e^{(-\beta \parallel r_{ij} \parallel ^2)}}
\end{equation}
where $\beta$ is the smoothing factor for the assignment. Soft-assignment assumes that different clusters have
equal scales. Inspired by Gaussian mixture models (GMM),
we further allow the smoothing factor $s_{k}$ for each cluster
center $c_{k}$ to be learnable:

\begin{equation}
    a_{ik} = \frac{e^{(-c_{k} \parallel r_{ik} \parallel ^2)}}{\sum_{j=1}^{K} e^{(-c_{j} \parallel r_{ij} \parallel ^2)}}
\end{equation}
which provides a finer modeling of the descriptor distributions. The Encoding Layer concatenates the aggregated residual vectors with assigning weights (as in Equation \ref{eq_1}). As is typical in prior work [1, 32], the resulting vectors are normalized using the L2-norm.

Now, all the components are differentiable with respect to (w.r.t) the input $X$ and the codewords $C = \{c_1, ...c_K\}$ and smoothing factors $s = \{s_1, ...s_k\}$. Therefore, the Encoding Layer can be trained end-to-end by standard SGD (stochastic gradient descent) with backpropagation.

\textbf{Gradients w.r.t Input $X$:} The encoder $E = \{\epsilon_1, ...\epsilon_K\}$ can be viewed as $k$ independent sub-encoders. Therefore the gradients of the loss function $\ell$ w.r.t input descriptor $x_i$ can be accumulated $\frac{d_\ell}{d_{x_i}} = \sum^K_{k=1} \frac{d_\ell}{d_{\epsilon_k}}.\frac{d_{\epsilon_k}}{d_{x_i}}$. According to the chain rule, the gradients of the encoder w.r.t the input is given by

\begin{equation}
    \frac{d_{\epsilon_k}}{d_{x_i}} = r^{T}_{ik}\frac{d_{a_{ik}}}{d_{x_i}} 
    + a_{ik}\frac{d_{r_{ik}}}{d_{x_i}} 
\end{equation}

where $\frac{d_{r_{ik}}}{d_{x_i}}=1$. Let $f_{ik}=e^{-sk\parallel r_{ik} \parallel^2}$ and $h_{i}=\sum^K_{m=1}f_{im}$, then $a_{ik}=\frac{f_{ik}}{h_i}$. The derivatives of the assigning weight w.r.t the input descriptor is

\begin{equation}
    \frac{d_{a_{ik}}}{d_{x_i}} = \frac{1}{h_i} . \frac{d_{f_{ik}}}{d_{x_i}} - \frac{f_{ik}}{(h_i)^2} .
    \sum^K_{m=1} \frac{d_{f_{im}}}{d_{x_i}}
\end{equation}

where $\frac{d_{f_{im}}}{d_{x_i}}=-2s_kf_{ik}.r_{ik}$.

\textbf{Gradients w.r.t Codewords $C$:} The sub-encoder $\epsilon_k$ only depends on the codeword $c_k$. Therefore, the gradient of loss
function w.r.t the codeword is given by $\frac{d_l}{d_{c_k}} = \frac{d_l}{d_{\epsilon_k}}. \frac{d_{\epsilon_k}}{d_{c_k}}$.

\begin{equation}
    \frac{d_{\epsilon_{k}}}{d_{c_k}} = \sum^N_{i=1}(r^T_{ik} \frac{d_{a_{ik}}}{d_{c_{k}}} + a_{ik} \frac{d_{r_{ik}}}{d_{c_{k}}})
\end{equation}

where $\frac{d_{r_{ik}}}{d_{c_{k}}}=-1$. Let $g_{ik} = \sum_{m\neq k} f_{im}$. According to the chain rule, the derivatives of assigning w.r.t the codewords can be written as

\begin{equation}
    \frac{d_{a_{ik}}}{d_{c_k}} = 
    \frac{d_{a_{ik}}}{d_{f_{ik}}} .
    \frac{d_{f_{ik}}}{d_{c_{k}}} =
    \frac{2s_k f_{ik} g_{ik}}{(h_i)^2} . r_{ik}
\end{equation}

\textbf{Gradients w.r.t Smoothing Factors:} Similar to the codewords, the sub-encoder $\epsilon_k$ only depends on the k-th smoothing factor $s_k$. Then, the gradient of the loss function w.r.t the smoothing weight is given by $\frac{d_l}{d_{s_k}} = \frac{d_l}{d_{\epsilon_k}}. \frac{d_{\epsilon_k}}{d_{s_k}}$.

\begin{equation}
    \frac{d_{\epsilon_{k}}}{d_{s_k}} = -
    \frac{f_{ik} g_{ik} \parallel r_{ik} \parallel ^2}{(h_i)^2}
\end{equation}

and in practice, we multiply the numerator and denominator of the assigning weight with $\epsilon^{\phi_i}$ to avoid overflow:

\begin{equation}
    a_{ik} = \frac{e^{(-s_k \parallel r_{ik} \parallel^2 + \phi_i)}}
    {\sum^K_{j=1} e^{(-s_j \parallel r_{ij} \parallel^2 + \phi_i)}}
\end{equation}
where $\phi_i=min_k\{s_k \parallel r_{ik} \parallel^2\}$. Then $\frac{d_{\overline{f}_{ik}}}{d_{x_i}} = e^{\phi_i} \frac{d_{f_{ik}}}{d_{x_i}}$.

%and the resulting vectors are normalized using the L2-norm:

In order to network learn patterns from different resolutions of images, the results of the aggregation for dictionaries of different layers are combined together (refer to Figure \ref{fig_MRDL}). However, depend on the task, importance of spatial texture in different information levels may not be the same and also varies from one task/data to another. Contribution of each dictionary in the model is determined by the weights ($\omega_{i}$) that are being learnt during the training (see Figure \ref{fig_MRDL}), which $\sum_{l=1}^{L} \omega_{l}=1$, and $0<\omega_{l}<1$. This gives the MRDL network the ability not only to dynamically and adaptively learn and adjust the importance of different layers, but also incorporate the effect of different texture levels by propagating the information to the loss function.

\section{Experimental Results and Discussion}
\label{Experiments}

In this section, we first describe the datasets that are used to carry out the experiments, and then report the experimental settings and present the results. 

\subsection{Datasets}
Four different histopathology image datasets consisting of digitized images of H\&E-stained glass slides of cancerous tissue specimens with different characteristics are used to investigate the performance of the proposed deep MRDL. Table \ref{table_datasets} briefly lists the main characteristics of each dataset.

\begin{figure}
\centering
\includegraphics[scale=0.45]{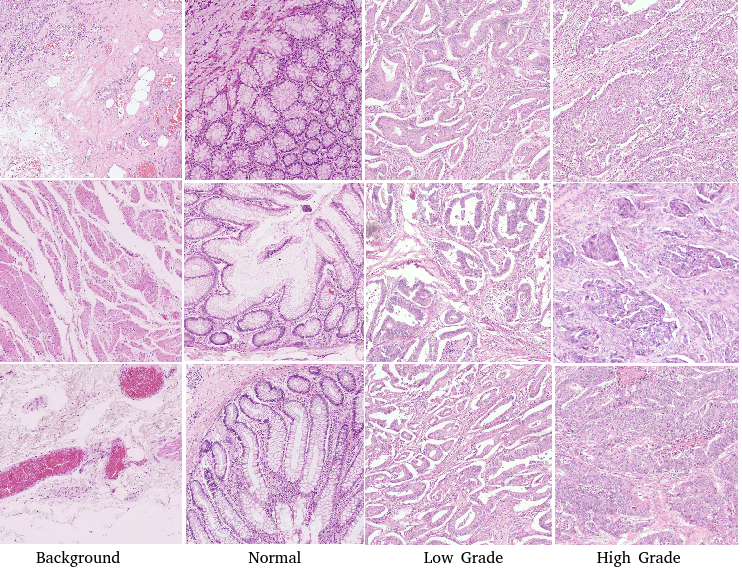}
\caption{Samples from 4 classes of the Colorectal Cancer Grading Dataset. First, second, third, and forth columns show samples from background, normal, low and high grade patches, respectively.}
\label{fig_crc-tia}
\end{figure}

\begin{figure}[!t]
\centering
\includegraphics[scale=0.43]{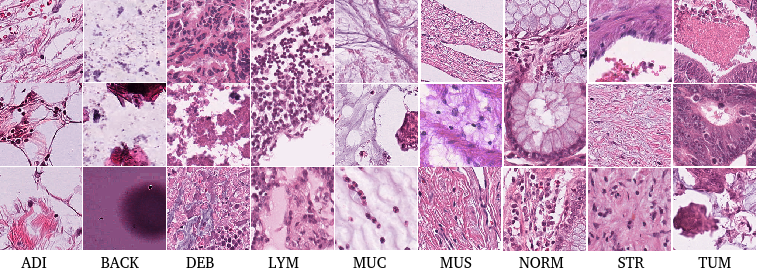}
\caption{Samples from the NCT-CRC-HE-100k dataset. 9 tissue classes are: Adipose (ADI), background (BACK), debris (DEB), lymphocytes (LYM), mucus (MUC), smooth muscle (MUS), normal colon mucosa (NORM), cancer-associated stroma (STR), colorectal adenocarcinoma epithelium (TUM).} 
\label{fig_kather100k}
\end{figure}

\begin{table}
\renewcommand{\arraystretch}{1.3}
\caption{Datasets used in our study. The table lists characteristics of the datasets and the total numbers of training, validation, and test images.}
\label{table_datasets}
\centering
\small\addtolength{\tabcolsep}{-5pt}
\begin{tabular}{|c|c|c|c|c|}
\hline
Dataset & CRC-TIA & PCam & NCT-CRC & BreakHis \\
\hline
\# of patches & 36,750 & 327,680 & 107,180 &  7,909\\
dimension & 1792 px$^{2}$ & 96 px$^{2}$ & 224  px$^{2}$& 700x460  \\
\# of slides  & 139 & 400 & 136 & 82 \\
\# of classes & 4 & 2 & 9  & 2 \\
magnification & 20x & 10x & 20x & 40x \\
organ & colorectal & lymph node & colorectal & breast\\
problem & grading & non/tumor & tissue type & benign/malign\\

\hline
\end{tabular}
\end{table}

\textbf{CRC-TIA} \cite{CRC_TIA1,CRC_TIA2}:
Grading of CRC tissues is a routine part of the pathological analysis and plays a key role in the treatment plan. %There are different grading systems for some cancers. 
Although there is no accepted standard grading system, many cancers use the following: \emph{Grade I} or low-grade that cancer cells resemble normal cells and usually grow more slowly, \emph{Grade II} or moderate that look more abnormal and are slightly faster growing, and \emph{Grade III} or high-grade look very different from normal cells and may grow more quickly. CRC-TIA dataset comprises of 36,750 non-overlapping images of size 1,750$\times$1,750 pixels, extracted at magnification 20$\times$. Each image is labelled as normal, low grade tumor or high grade tumor by an expert pathologist. To obtain these images, digitized WSIs of 139 CRC tissue slides stained with H\&E are used. All WSIs were taken from different patients and were scanned using the Omnyx VL120 scanner at 0.275 $\mu$m/pixel (40$\times$). Samples of the dataset are shown in Figure \ref{fig_crc-tia}. Each image in the CRC-TIA dataset consists of $n$ tissue patches. In this paper, the final image-level classification label is obtained by aggregating the patch-level labels using majority voting.

\begin{figure}[!t]
\centering
\includegraphics[scale=0.205]{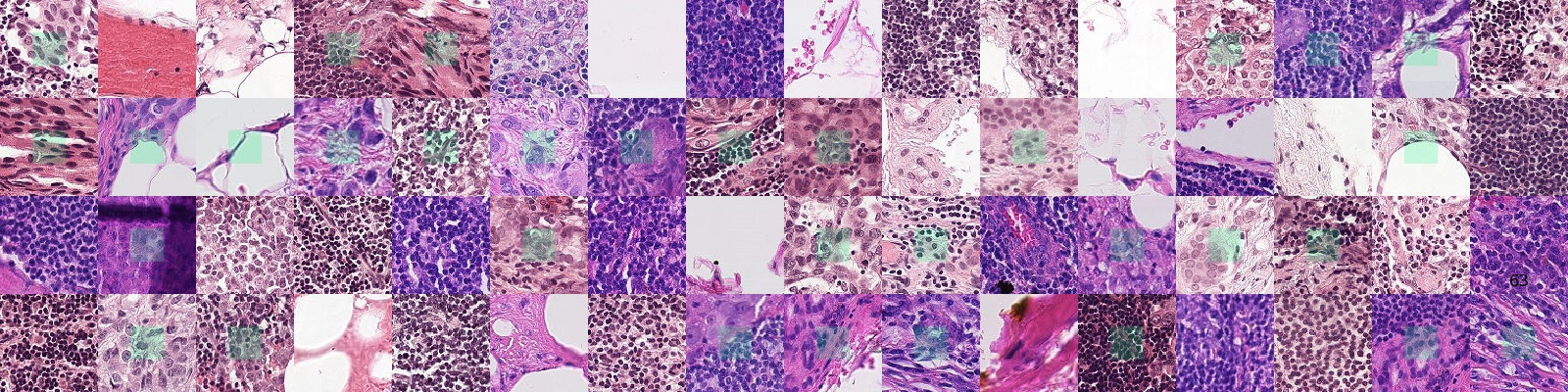}
\caption{Examples from PCam data. Green boxes indicate tumor images.} 
\label{fig_PCam}
\end{figure}

\textbf{PCam} \cite{PCam}: The PatchCamelyon metastasis detection benchmark consists of 327,680 color images (96$\times$96 pixels) extracted from digital scans of breast cancer lymph node sections. 
PCam is derived from the Camelyon16 Challenge \cite{bejnordi2017diagnostic}, which contains 400 H\&E stained WSIs of sentinel lymph node sections. The slides were acquired and digitized at 2 different centers using a 40$\times$ objective. We then undersample images at 10$\times$ to increase the field of view. We follow the train/test split from the Camelyon16 challenge, and further hold-out 20\% of the train WSIs for the validation set. To prevent selecting background patches, slides are converted to HSV, blurred, and patches filtered out if maximum pixel saturation lies below 0.07. Each image is given a binary label indicating presence of metastatic tissue. Samples of the dataset are shown in Figure \ref{fig_PCam}.

\textbf{NCT-CRC} \cite{kather100_data}:
The training data (NCT-CRC-HE-100K) is a set of 100,000 non-overlapping image patches from H\&E-stained histological images of human colorectal cancer (CRC) and normal tissue. All images are 224$\times$224 pixels at 0.5 microns per pixel (MPP). All images are color-normalized using Macenko's method \cite{macenko2009}. Tissue classes are: Adipose (ADI), background (BACK), debris (DEB), lymphocytes (LYM), mucus (MUC), smooth muscle (MUS), normal colon mucosa (NORM), cancer-associated stroma (STR) and colorectal adenocarcinoma epithelium (TUM). The images were manually extracted from 86 H\&E stained cancer tissue slides from formalin-fixed paraffin-embedded (FFPE) samples from the NCT Biobank (National Center for Tumor Diseases, Heidelberg, Germany) and the UMM pathology archive (University Medical Center Mannheim, Mannheim, Germany). Tissue samples contained CRC primary tumor slides and tumor tissue from CRC liver metastases; normal tissue classes were augmented with non-tumorous regions from gastrectomy specimen to increase variability. The validation data (NCT-CRC-HE-7K) is a set of 7,180 image patches from N=50 patients with colorectal adenocarcinoma (no overlap with patients in NCT-CRC-HE-100K). Samples of the dataset are shown in Figure \ref{fig_kather100k}.

\textbf{BreakHis} \cite{spanhol2015dataset}:
The Breast Cancer Histopathological Image Classification (BreakHis) is composed of 9,109 microscopic images of breast tumor tissue collected from 82 patients using different magnifying factors (40$\times$, 100$\times$, 200$\times$, and 400$\times$). It contains 2,480 benign and 5,429 malignant samples (700$\times$460 pixels, 3-channel RGB, 8-bit depth in each channel, PNG format). This database has been built in collaboration with the P\&D Laboratory – Pathological Anatomy and Cytopathology, Parana, Brazil. 

\subsection{Settings and Results}
We carried out experiments in TensorFlow.Keras using Python. Seven popular CNN models VGG16 \cite{VGG16}, Xception \cite{Xception}, ResNet18 \cite{ResNet}, ResNet50 \cite{ResNet}, MobileNet \cite{MobileNet}, InceptionV3 \cite{szegedy2015going}, and DenseNet \cite{DenseNet} with pre-trained ImageNet weights are used as baselines and also as backbones for extracting deep descriptors for the MSDL. In order to obtain the best baseline results, SGD and Adam optimizers with learning rates of 0.01 and 0.001 are tried with number of epochs 20 and batch size of 64. Horizontal and vertical flips and rotation range of 40 degrees are applied for data augmentation. For MSDL, features from ``global\_average\_pooling2d'' layer are pooled from each model. Regardless of backbone model, the MSDL has two main parameters to set: the size of dictionary ($D$) and the dimension of descriptors (N). In our experiments, we explored $D:\{8,16,32,64,128\}$ and $N:\{32,64,128,256,512\}$, and the best $\{D,N\}$ is selected based on cross-validation performance.

\begin{table}
\renewcommand{\arraystretch}{1.3}
\caption{MRDL performance (in terms of accuracy \%) using different popular CNN backbones.}
\label{table_diff_backbones}
\centering
\begin{tabular}{|c|c|c|c|c|}
\hline
Backbone & CRC-TIA & PCam & NCT-CRC & BreakHis \\
\hline
ResNet18 \cite{ResNet} & 92.2  & \bf{90.4} & 96.0 & 95.4 \\
ResNet50 \cite{ResNet} & 92.0 & 90.0 & 95.9 & 95.4 \\
VGG16 \cite{VGG16} & 90.6 &  88.9 & 96.2 & 95.0 \\
MobileNets \cite{MobileNet} & 89.0 &  89.4 & 95.5 & 96.1 \\
DenseNet \cite{DenseNet} & 91.8 & 90.2 & 96.1 & \bf{96.9} \\
InceptionV3 \cite{szegedy2015going} & 91.4 & 90.0 & 96.1 & 96.4 \\
Xception \cite{Xception} & \bf{92.5} & 89.7 & \bf{96.6} & 95.0 \\

\hline
\end{tabular}
\end{table}

Effect of different popular CNN backbones are investigated in Table \ref{table_diff_backbones}. As shown, there is no single CNN models among VGG16, Xception, ResNet18, ResNet50, MobileNet, InceptionV3, and DenseNet that outperform the others across all datasets. However, all the backbones demonstrate similar performance. In our experiments, Xception for CRC-TIA, ResNet18 for PCam, Xception for NCT-CRC, and DenseNet for BreakHis are obtained a marginal superior accuracies, respectively.

\begin{table}
\renewcommand{\arraystretch}{1.3}
%\extrarowheight as needed to properly center the text within the cells
\caption{Effect of different dictionary sizes on the classification performance (in terms of accuracy \%).}
\label{table_diff_D}
\centering
\begin{tabular}{|c|c|c|c|c|}
\hline
Dataset & D=16 & D=32 & D=64  & D=128\\
\hline
CRC-TIA & 90.9  & 92.0 & \bf{92.5} & 91.9 \\
PCam & 89.3 & 89.1 & \bf{90.4}  & \bf{90.4} \\
NCT-CRC & 95.9 &  \bf{96.6} & 96.1 & 96.0 \\
BreakHis & 96.2 &  \bf{96.9} &  96.2 & \bf{96.9} \\
\hline
\end{tabular}
\end{table}

First sets of experiments are dedicated to investigate the effect of different parameters and sensitivity of the MRDL model respect to them. One of the main parameters of MRDL network is dictionary size $D$. From traditional BoW model, it is know that too small dictionary has a less capacity to observe the data distribution and demonstrates limited ability to learn the task at hand properly. However, too big dictionary size than needed will also overfit on the training data and will be too noisy. Therefore, finding the optimal $D$ is desirable for each task and selected backbone networks. Table \ref{table_diff_D} reports the effect of different $D$s on performance (in terms of accuracy \%) of MRDL. As expected, there is no universal winner $D$ for all datasets. However, it is obvious that $D=16$ is too small for all datasets and has a limited capacity to properly learn the problem. Since there is no obvious winner, $D=64$ seems to be a good size and will be used for the rest of this study.

Another interesting experiments is to investigate the contribution of each hierarchy layer on MRDL. Multi-resolution analysis is presented in Table \ref{table_diff_L}. Effect of different dictionaries of different resolutions on the classification performance in reported while keeping the $D$ fixed at size 64. Three different results are reported for each dataset. Network with $l=3$ is a network with only one dictionary generated from the last layer of the backbone network. This dictionary only uses the last-level information as descriptors and the network can not be considered as multi-resolution. Network with $l=2,3$ has two dictionaries, one from the last layer, and the other from previous spatial resolutions. And finally network with $l=1,2,3$ contains three dictionaries from three different spatial resolutions as shown in Figure \ref{fig_MRDL}. Different individual dictionaries are weighted summed before being fed into FC layers. As expected and shown in Table \ref{table_diff_L}, for all four datasets the performance is evolving as more spatial resolution is added to the network. This proofs the hypothesis that dictionaries of different hierarchy level contain different information that positively incorporate in entire network.

\begin{table}
\renewcommand{\arraystretch}{1.3}
%\extrarowheight as needed to properly center the text within the cells
\caption{Multi-resolution analysis: Effect of different dictionaries of different resolutions on the classification performance (in terms of accuracy \%).}
\label{table_diff_L}
\centering
\begin{tabular}{|c|c|c|c|}
\hline
Dataset & $l$=3 & $l$=2,3 & $l$=1,2,3  \\
\hline
CRC-TIA & 92.2  & 92.2 & \bf{92.5}  \\
PCam & 89.2 & 89.7 & \bf{90.4}  \\
NCT-CRC & 95.9 &  96.0 & \bf{96.6}  \\
BreakHis & 95.8 &  96.4 & \bf{96.9}  \\
\hline
\end{tabular}
\end{table}

\begin{figure}
    %\begin{subfigure}
        \includegraphics[width=0.4\textwidth]{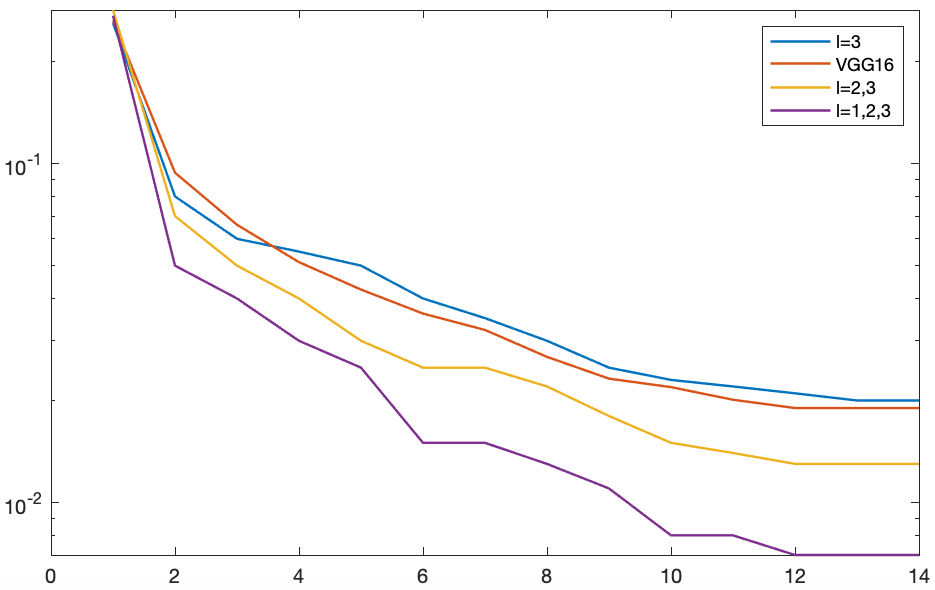}
    %\end{subfigure}
    %\begin{subfigure}
        \includegraphics[width=0.4\textwidth]{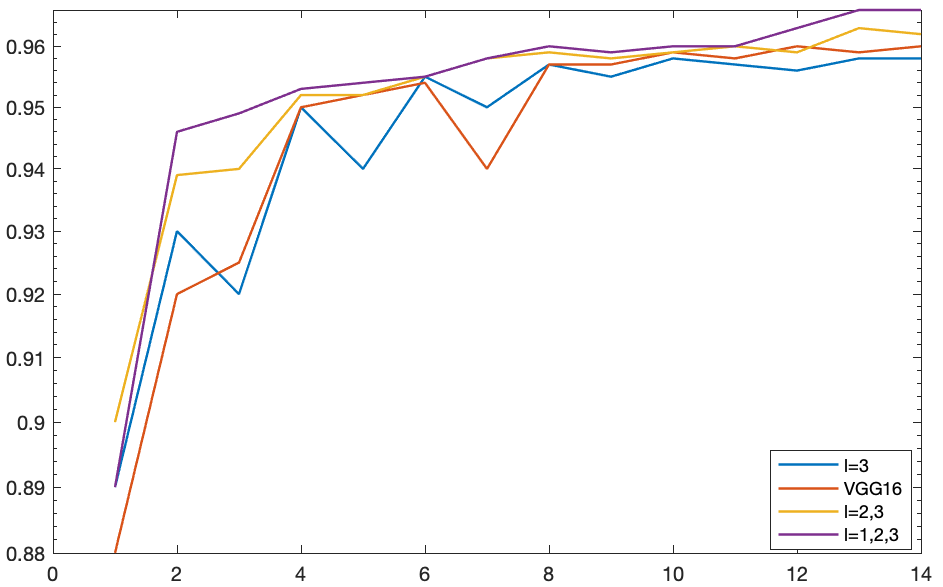}
    %\end{subfigure}
    \caption{Plots of learning curves on NCT-CRC dataset. The training losses (top), and validation accuracy (bottom) of an exemplar MRDL with different resolutions $l$ compared to its VGG16 backbone.} 
\label{fig_loss_acc_kather100k}
\end{figure}

Plots of learning curves on NCT-CRC dataset is shown in Figure \ref{fig_loss_acc_kather100k}. The training losses (top), and the validation accuracy (bottom) of an exemplar MRDL with different resolutions $l$ is presented and compared to its VGG16 backbone. The training loss of standard VGG16 is presented as a reference in order to compare how well MRDL model performs, and also to see the effect of adding different spatial resolutions on optimization of the networks during the epochs. The Y-axis is in log-scale for a better visualization, and the X-axis represents 14 epochs until the networks converge. Comparing four different models, one can observe that the performance of deep dictionary network without additional resolutions ($l=3$) is slightly better than its VGG16 backbone. The second observation is that introducing more spatial resolution reduces the training loss and improves the validation accuracy.

%\begin{figure}
%\centering
%\includegraphics[scale=0.2]{loss_kather100.png}
%\caption{Plot of learning curves on NCT-CRC dataset. The training losses (top), and validation accuracy (bottom) of an exemplar MRDL with different resolutions $l$ compared to its VGG16 backbone.} 
%\label{fig_loss_kather100k}
%\end{figure}

\begin{table}[!t]
\renewcommand{\arraystretch}{1.3}
\caption{Comparison with state-of-the-art on CRC-TIA dataset.}
\label{table_crc_state_of}
\centering
\begin{tabular}{|c|c|c|}
\hline
Method &  Patch Accuracy & Image Accuracy  \\
\hline
BAM-1 \cite{CRC_TIA1} & 87.79 & - \\
Context-G \cite{sirinukunwattana2018improving} &89.96 &-  \\
BAM-2 \cite{CRC_TIA1} & 90.66 & -  \\
InceptionV3 \cite{szegedy2015going} & 91.21 & 96.20 \\
ResNet50 \cite{ResNet} & 92.23 &  96.44 \\
Xception \cite{Xception} & 92.19 & 96.86 \\
MobileNet \cite{MobileNet} & 92.44 & 97.25  \\
CA-CNN \cite{Shaban20} & 95.70 & - \\
CGC-Net \cite{Zhou19} & 91.60 & 97.00  \\
\bf{deep MRDL}   &  \bf{92.50}& \bf{97.73}\\

\hline
\end{tabular}
\end{table}

Performance of the proposed deep MRDL is compared to the state-of-the art results. Tables \ref{table_crc_state_of}, \ref{table_PCam_state_of} and \ref{table_BreakHis_state_of}. Although there are some traditional shallow algorithms among the state of the art (e.g. BAM \cite{CRC_TIA1}), the majority of them are deep learning-based approached. Please note that the MRDL main parameters i.e. $D$, $N$ and also backbone type is optimized for each dataset, and varies from one another. The results are reported in terms of the classification accuracies for all the tasks. Additionally, the area under the curve (AUC) is also provided for \ref{table_PCam_state_of}. As shown, the performance of the deep MRDL outperforms the state-of-the art results across all the datasets.

\begin{table}[!t]
\renewcommand{\arraystretch}{1.3}
\caption{Comparison with state-of-the-art on PCam dataset.}
\label{table_PCam_state_of}
\centering
\begin{tabular}{|c|c|c|}
\hline
Method &  Accuracy & AUC  \\
\hline

InceptionV3 \cite{szegedy2015going} & 87.5 & 95.1 \\
Xception \cite{Xception} & 88.3 & 94.9 \\
MobileNet \cite{MobileNet} & 89.0 & 95.7  \\

ResNet18 \cite{ResNet} & 88.7 & 95.3 \\
DenseNet \cite{DenseNet} &  87.6 & 95.5 \\
GDensenet \cite{PCam} & 89.8 & 96.3  \\
\bf{deep MRDL}   &  \bf{90.4} & \bf{96.8}\\

\hline
\end{tabular}
\end{table}

\begin{table}[!t]
\renewcommand{\arraystretch}{1.3}
\caption{Comparison with state-of-the-art on BreakHis dataset.}
\label{table_BreakHis_state_of}
\centering
\begin{tabular}{|c|c|}
\hline
Method &  Accuracy  \\
\hline
Spanhol {\em et al.} \cite{spanhol2017deep}  & 84.2\\
Spanhol {\em et al.} \cite{spanhol2016breast}  & 84.6\\
DenseNet \cite{DenseNet} &  93.5 \\
GDensenet \cite{PCam} & 96.1  \\
Vo {\em et al.} \cite{vo2019classification} & 96.3 \\
\bf{deep MRDL}   &  \bf{96.9}\\

\hline
\end{tabular}
\end{table}

\begin{figure*}[!t]
\includegraphics[width=\textwidth]{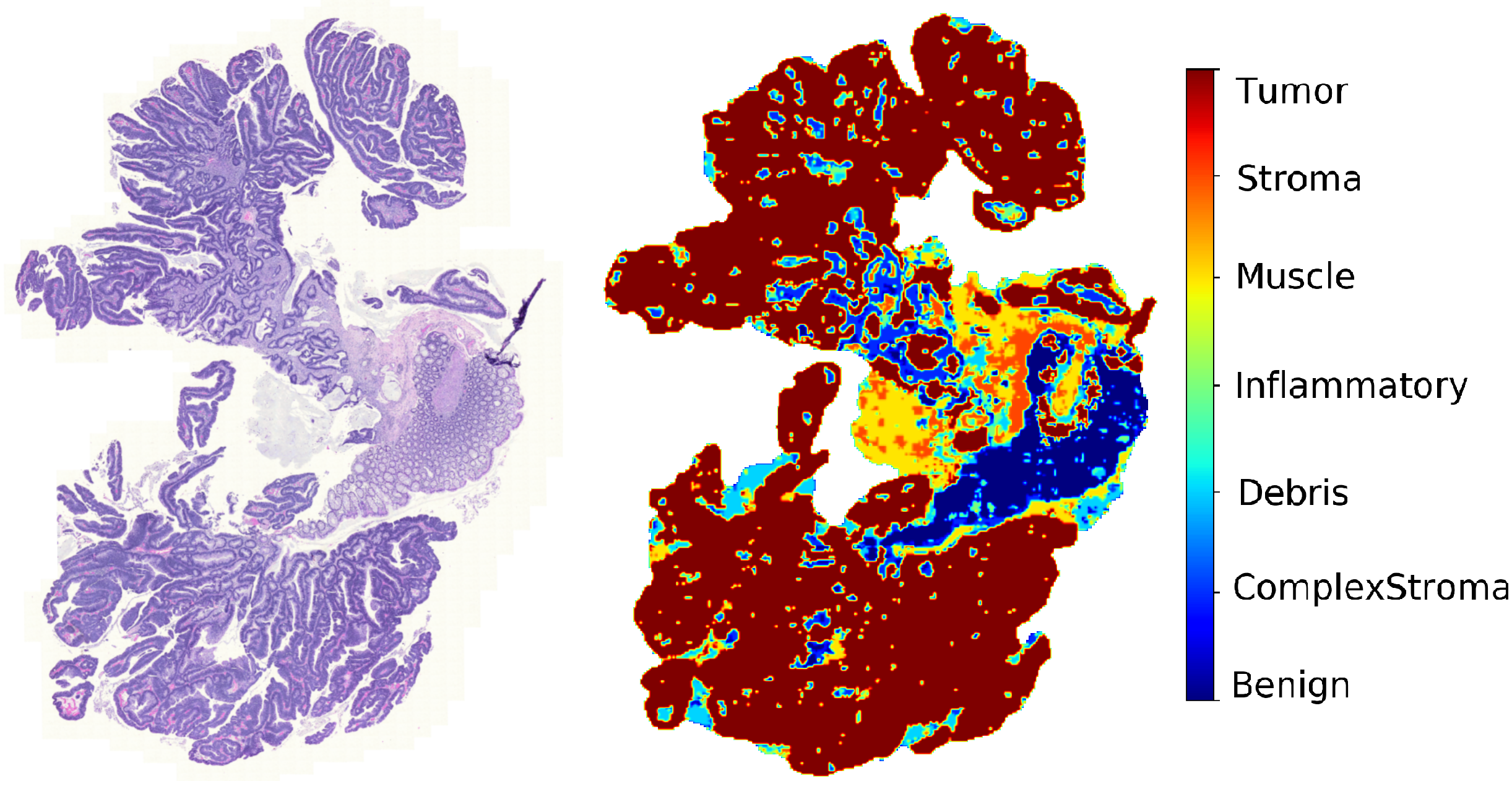}
\caption{Qualitative Evaluation: The results of the proposed deep MRDL algorithm for tissue phenotyping on an unseen test WSI taken from TCGA repository. Left: the original colorectal WSI; Right: tissue phenotyping result. The classifier is trained on the NCT-CRC dataset \cite{kather100_data}. (The figure is best seen in color).}
\label{fig_pt_00}
\end{figure*}

\begin{figure*}
\includegraphics[width=\textwidth]{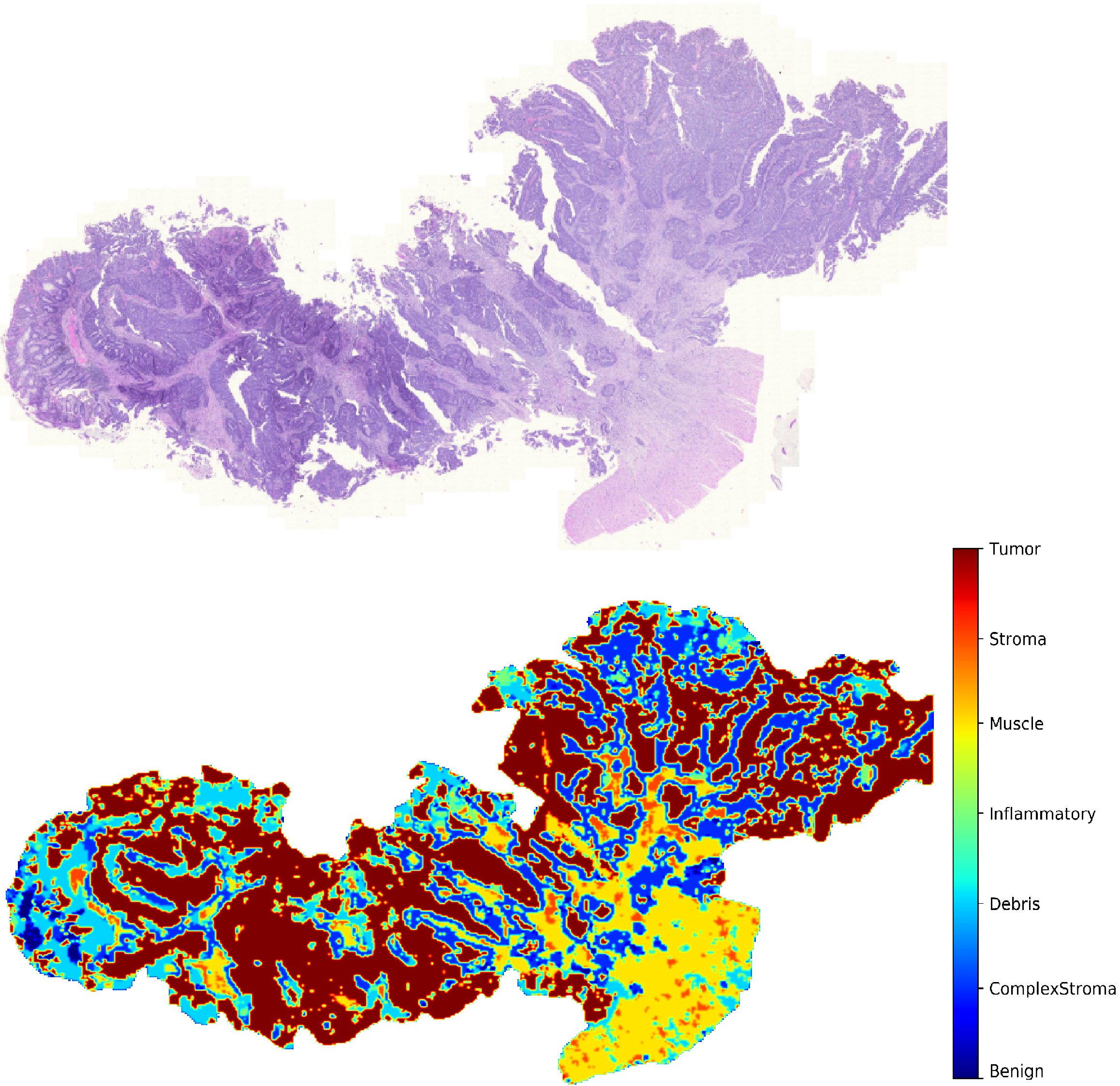}
\caption{Qualitative Evaluation: The results of the proposed deep MRDL algorithm for tissue phenotyping on an unseen test WSI taken from TCGA repository. Top: the original colorectal WSI; Bottom: tissue phenotyping result. The classifier is trained on the NCT-CRC dataset \cite{kather100_data}. (The figure is best seen in color).}
\label{fig_pt_01}
\end{figure*}

Figure \ref{fig_pt_00} and \ref{fig_pt_01} show the deep MRDL output for automatic phenotyping of an exemplar Colon Adenocarcinoma (COAD) slides from the TCGA repository. The classifier is trained on the NCT-CRC dataset \cite{kather100_data} with seven classes of tumor, stroma, muscle, inflammatory, debris, complex stroma, and benign tissues. 
Afterwards, it tested on $224\times 224$ non-overlapping patches of an unseen WSI (patient-level separation). These patches are selected from the tissue regions after applying Otsu-based tissue segmentation algorithm on the WSI at $5\times$. The color normalization is applied on the patches before inferring, and smoothing of the predicted labels is applied using median filter to remove the blocky effects after constructing the WSI. In Figure \ref{fig_pt_00}, left image is the original $12352\times 17984$ pixels WSI at $20\times$ and the right is heatmap output. Tumor regions are shown with red color, stroma, muscle, inflammatory, debris, complex stroma and benign regions are shown with orange, yellow, green, turquoise, light blue, and dark blue respectively. Similarly, In Figure \ref{fig_pt_01}, top image is the original $19392 \times 24128$ pixels WSI at $20\times$ and the bottom is heatmap output, with color labels similar to previous figure \ref{fig_pt_00}. In both figures, the resulting colormaps are manually examined by experienced pathologist and found to be matching with manual assessment.

\section{Conclusions}
\label{Conclusions}

Application of deep dictionary learning for histopathology image analysis is investigated. In order to incorporate deep descriptors containing spatial information of different layers, MRDL model is proposed. A separate dictionary is generated for each information level at each spatial layer, and the dictionaries are adaptively and dynamically combined taking into account their contribution. Experiments are carried out on different computational pathology tasks such as colorectal cancer grading, colorectal tissue phenotyping, lymph node tumor detection, breast malignancy detection using different CNN backbones. Also, effect of different dictionary sizes and dictionaries of different spatial resolutions on the classification performance is studies. Promising results demonstrate the potential of the proposed deep dictionary learning, specially MRDL for computational pathology.

One future direction could be adaptation of the MRDL on giga-pixel WSIs. Instead of applying MRDL on a patch-level, it would be desirable to train and infer the MRDL on WSI at once. Application of deep dictionary learning, specially the proposed deep MRDL on patient output, e.g. survival prediction is also another interesting research direction.

% or


\section*{Acknowledgments}

This work was supported by the UK Medical Research Council (award MR/P015476/1). NR is also supported by the PathLAKE digital pathology consortium, which is funded from the Data to Early Diagnosis and Precision Medicine strand of the government’s Industrial Strategy Challenge Fund, managed and delivered by UK Research and Innovation (UKRI).


%\ifCLASSOPTIONcaptionsoff
%  \newpage
%\fi



%\bibliographystyle{IEEEtran}

%\begin{thebibliography}{1}
%\end{thebibliography}

\bibliographystyle{IEEEtran}
\bibliography{refs.bib}

%\begin{IEEEbiography}[{\includegraphics[width=1in,height=1.25in,clip,keepaspectratio]{a1.png}}]{Nima Hatami} ...
%\end{IEEEbiography}

%\begin{IEEEbiography}[{\includegraphics[width=1in,height=1.25in,clip,keepaspectratio]{a2.png}}]{Mohsin Bilal} ...

%\end{IEEEbiography}

%\begin{IEEEbiography}[{\includegraphics[width=1in,height=1.25in,clip,keepaspectratio]{a3.png}}]{Nasir Rajpoot} ...
%\end{IEEEbiography}

\end{document}